\documentclass[prl,twocolumn,showpacs,preprintnumbers,amsmath,amssymb]{revtex4}

\usepackage{graphicx}
\usepackage{dcolumn}
\usepackage{bm}

\begin{document}

\preprint{APS/123-QED}

\title{Suppression of Reabsorption via Modulation of Light}

\author{Anthony R. Gorges, Ansel J. Foxley, David M. French, Christopher M. Ryan, and Jacob L. Roberts}
 
\affiliation{%
Department of Physics, Colorado State University, Fort Collins, CO 80523
}%

\date{\today}

\begin{abstract}
Reabsorption, the multiple scattering of spontaneously emitted photons in optically thick gases, is a major limitation to efficient optical pumping and laser cooling in ultracold gases.  We report mitigation of reabsorption using spatial and frequency modulation of laser light illuminating such gases.  We developed a semi-classical model that successfully describes the reabsorption process when frequency-modulated light is present.  It was necessary to extend the treatment in the model beyond a simple two-atom picture in order to reproduce our experimental results.
\end{abstract}

\maketitle

Reabsorption has continually plagued laser cooling schemes by critically limiting the efficiency of optical pumping, and therefore cooling rates, in optically thick gases of ultracold atoms.  This is because the multiple scattering of spontaneously emitted photons in these gases, which constitutes reabsorption, results in undesirable heating and depolarization of the optically pumped atoms.  The limitations due to reabsorption have been observed experimentally in studies showing a decrease in laser cooling efficiency with increasing optically thickness \cite{Weiss2001,Chu1998,Salomon1999,Chu2000,Salomon1996}.  The effects of reabsorption are believed to be a main reason why Bose-Einstein Condensates (BECs) cannot be created using non-evaporative laser cooling techniques.  Understanding and mitigating the effects of reabsorption should allow marked improvements to optical pumping and laser cooling in ultracold gases.

\begin{figure}
\includegraphics{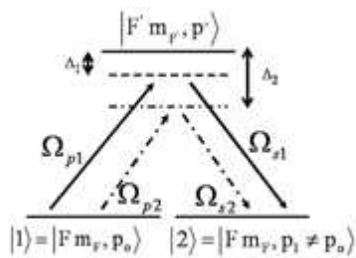}
\caption{\label{label}A picture of reabsorption. The quantum states (solid horizontal lines) and light (diagonal lines) involved in the two-photon reabsorption process are illustrated.  The atom is driven from one momentum state ($p_o$) to another ($p_1$) in the same hyperfine magnetic sublevel by a combination of pump light ($\Omega_{p1}, \Omega_{p2}$) and spontaneous emission in the ultracold cloud ($\Omega_{s1}, \Omega_{s2}$).  While reabsorption still occurs when a light with a single frequency illuminates the cloud, in this figure we illustrate the case in which two lasers with two different detunings ($\Delta_1$ and $\Delta_2$) are used.}
\end{figure}
To date, experimental and theoretical work \cite{Lewenstein1998,Zoller,Weiss2000,Prentiss,Lewenstein2001} has focused on the use of trap geometry and tightly confining trap potentials to decrease reabsorption.  For instance, by using elongated traps the average density of the gas can be maintained while reducing the optical depth in one direction out of the cloud, making it easier for photons to escape out of the narrow direction \cite{Lewenstein1998,Prentiss}.  In other work, optical lattices have been used to mitigate reabsorption since their naturally strong confinement potentials allow for operation in the $\it{festina}$ $\it{lente}$ regime\cite{Weiss2000,Chu1998}.  

In contrast, little attention has been paid to a careful consideration of how alterations of the optical pumping or laser cooling light could be used to reduce reabsorption effects.  This Letter presents experimental evidence of the reduction of reabsorption by simultaneously altering the frequency spectrum of and creating a spatial interference pattern in the applied laser light.  These techniques are broad reaching and should reduce the amount of reabsorption regardless of trap geometry or confinement strength.  Furthermore, this paper presents a semi-classical model which explains our experimental results.  Unlike previous theoretical descriptions \cite{Lewenstein1998,Zoller}, our model includes physics beyond a simple two-atom treatment since a more complete description of the nature of spontaneous emission and interparticle correlations in a many-atom ultracold gas needs to be taken into account in order to explain our experimental observations.

There are two reabsorption processes, both of which rely on pump and scattered photons \cite{photons}.  The first process is a one-photon scattering wherein a newly created scattered photon scatters one or more times before escaping the optically thick gas.  While this multiple scattering introduces undesirable heating and depolarization of the atoms, the one-photon process is dependent upon the detuning of the laser and thus can be nearly eliminated with a sufficiently large detuning.  The second process is a two-photon Raman scattering phenomenon in which a scattered photon simultaneously scatters with a pump photon of the same frequency (Fig.\ 1).  Since the scattered photons do not in general travel in the same direction as the pump photon, this two-photon scattering drives atoms from one momentum state to another. The net effect is momentum diffusion that results in heating.  Unlike the one-photon process, the two-photon process is independent of detuning since scattered photons have the same frequency as the pump photons which create them, meaning the two-photon process is ultimately the main contributor to reabsorption.  In the remainder of this Letter, reabsorption will therefore refer to the two-photon process.

Because reabsorption relies on the pump and scattered light being the same frequency, one could imagine that using two lasers with different detunings should reduce the amount of reabsorption, since not all of the pump and scattered photons will have the same frequency.  Guided by this reasoning, we investigated the effect the number of distinct frequency components in the pump light had on reabsorption by illuminating an ultracold cloud of atoms with pump light containing one or more frequency components and measuring the resulting heating.  Since the heat imparted is related to the amount of reabsorption, measuring the imparted heat allowed us to determine if alterations in the pump light altered the reabsorption.  We were able to reduce the amount of reabsorption with multiple-frequency pump light.  However, we found that using frequency modulation alone was not sufficient to reduce reabsorption; the beams associated with the different frequency components must create an interference pattern in space with a characteristic length scale smaller than the mean free path of the photons in the cloud. This was illustrated in the results of two different data sets, one with an interference pattern and one without, in which an ultracold cloud of $^{85}$Rb atoms was heated by laser light with two distinct frequency components.
\begin{figure}
\includegraphics{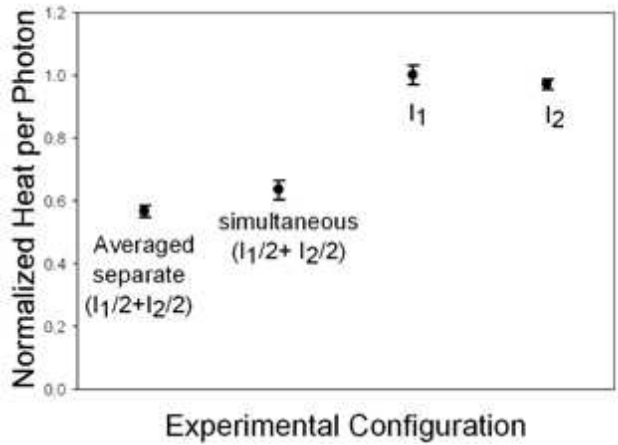}
\caption{\label{label}Normalized Heat per Photon vs Experimental Configuration for two laser experiment.  Conditions are labeled and clearly show a difference between two $I/2$ beams illuminating the cloud simultaneously and one beam at $I$.}
\end{figure}

To test the effect of multiple frequency component pump light on reabsorption, $^{85}$Rb atoms were first prepared in a magneto-optic trap (MOT) using standard techniques \cite{Pritchard}.  A typical experimental cycle involved the MOT being loaded, a 15 ms CMOT stage \cite{Cornell}, and then 3 ms of an optical molasses.  The MOT, at this point, had temperatures between 10 and 20 $\mu K$ with a peak optical depth (OD) of approximately 10 and atom numbers of about 100 million.  The cooling light was then shut off and 1 ms later the cloud was heated by one or more laser beam pulses.  An Acoustic Optical Modulator (AOM) controlled the heating pulses, with typical pulse lengths of 40-1600 $\mu$s.  The heat pulses were set to be in the range of 40 to 60 MHz blue detuned of the $^{85}$Rb cycling transition.  After the heating pulse, the atoms underwent an 18 ms free expansion before finally being imaged via absorption imaging.  From the size of the cloud after expansion, the kinetic energy increase of the atoms due to the heating pulses was determined.  Changes in the amount of reabsorption manifested themselves as a change in the amount of heat (kinetic energy) imparted by the heating pulses.

To characterize these changes, we introduce a quantity called "heat per photon".  Heat imparted in a cloud in the absence of saturation goes as $C_{1}I\Delta t$+$C_{2}$$I^2$$\Delta t$ where $C_{1}$ is a coefficient of one-photon scattering, $C_{2}$ is a coefficient of two-photon scattering, $I$ is the total intensity of the beam, and $\Delta t$ is the pulse length.  Heat per photon is defined as $C_{1}$+$C_{2}I$, the heat divided by $I \Delta t$.  This quantity allows us to compare data sets taken under different intensity and pulse length conditions.

\begin{figure}
\includegraphics{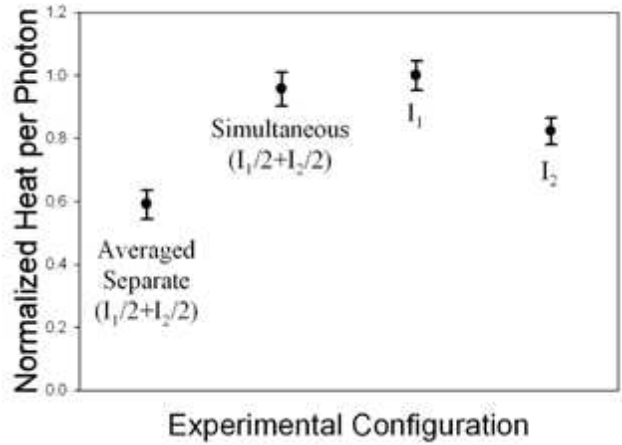}
\caption{\label{label}Normalized Heat per Photon vs Experimental Configuration for single laser/multiplexed AOM experiment.  Conditions are labeled and show no difference between two $I/2$ beams illuminating the cloud simultaneously and one beam at $I$.}
\end{figure}

Our first set of experiments used two separate lasers (laser 1 and laser 2) detuned by several MHz or more from each other to generate the heating pulses.  These pulses were co-propagated and used to simultaneously apply heat to the atom cloud.  The separate laser beams had different phase  and amplitude spatial profiles, and so overlapping the beams produced a spatial interference pattern.  The heating pulses had a 1.5 mm spot size and had an intensity $I/I_{sat}$ of about 10 at full intensity.  40 $\mu$s pulses were used to heat the cloud under four different conditions.  Three of these conditions were: laser 1 at $I_1$; laser 2 at $I_2$; and laser 1 at $I_1$/2 and laser 2 at $I_2$/2 simultaneously.  Since $I_1$$\sim$$I_2$, all three of these conditions had roughly the same total intensity.  Any reduction in reabsorption would be expected to result in the simultaneous data imparting less heat to the cloud than the full intensity conditions.  The fourth condition was an average of laser 1 alone at $I_{1}/2$ and laser 2 alone at $I_{2}/2$, called the averaged separate data. The averaged separate data represents a lower limit to the amount of reduction in reabsorption which could be realized.  Data were taken with a variety of intensities, detunings, atom number, and pulse lengths.  A sample set of data is shown in Fig.\ 2, but the same general trends were observed in all data sets.  There is no statistical difference between the averaged separate and simultaneous conditions.  The full intensity conditions, however, show a clear increase as compared to the simultaneous.

In the second set of experiments, a single laser was used to create two separate heating pulses.  This was accomplished by sending two radio frequencies into an AOM which produced two deflected, non-overlapped beams separated in frequency by 10 MHz.  The results of a typical experiment are shown in Fig.\ 3.  The conditions shown are the same as in the previous experiment, with the laser 1 and laser 2 now referring to the two AOM deflected beams.  Unlike the first set of experiments, no spatial interference pattern was present since the two beams were separate deflections of the same original beam.  In contrast to the first set of experiments, the heat per photon for the simultaneous case was similar to that of the full intensity individual beams.

To explain the dependence of our results on the presence of an interference pattern, we developed a model of the two-photon reabsorption process where the pump light and the spontaneous emission light are treated  semi-classically for a typical atom in the cloud.  The pump and spontaneous emission light are described by time-dependent Rabi frequencies $\Omega_p$ and $\Omega_s$, respectively. Once $\Omega_p$ and $\Omega_s$ are known, the transition rate from one momentum state to another (i.e from $\left|1 \right>$ to $\left|2 \right>$ in Fig.\ 1) is calculated to estimate heating rate due to reabsorption.

We assume that there are two distinct frequency components in the pump light and so we write (in a rotating-wave approximation) $\Omega_p=\Omega_1+\Omega_2e^{i \Delta \omega t}$, where $\Delta\omega$ is the difference between the frequencies of the two components and $\Omega_{\alpha}=\frac{d E_{\alpha}}{2\hbar}$, where $\alpha$=1 or 2, $d$ is the dipole matrix element, and $E_{\alpha}$ is equal to the electric field amplitude associated with beam $\alpha$.  To model the spontaneous emission light, we treat each atom as a classical oscillator driven by the pump light.  Thus, $\Omega_s=\kappa\sum^{N}_{i=1}e^{i\phi_i}(\Omega_1+\Omega_2e^{i\Delta\omega t+i\theta_i})$ where $\Omega_\alpha$ and $\Delta\omega$ have the same definitions as above and $\kappa$ is a constant \cite{kappanote}.  The summation is over the $N$ atoms whose spontaneous emission is in the proper direction to contribute to the $\left|1 \right>$ to $\left|2 \right>$ transition.  Because the spacing of the atoms in the gas is random and on average greater than the pump light wavelength, the phase of the spontaneous emission from each atom ($\phi_i$) is random as well.  $\theta_i$ allows for spatial variation in the phase between the two frequency components (i.e. an interference pattern).  $\Omega_s$ can be written in the form $\Omega_s=\kappa'(\Omega_1e^{i\phi'}+\Omega_2^{i\theta'})$ where $\kappa'$ will vary from atom to atom but will on average be $\sqrt{N}\kappa$.  Under our assumptions, $\phi'$ will vary randomly from atom to atom as well.  The relationship between $\theta'$ and $\phi'$ depends on the values of $\theta_i$.  If there is no spatial interference pattern, then $\theta_i=0$ for all $i$ and $\theta'=\phi'$.  In the opposite limit of a rapidly varying interference pattern, $\theta_i$ and thus $\theta'$ will be random, with $\theta'$ having no correlation to $\phi'$.

With this model of $\Omega_p$ and $\Omega_s$, we can calculate the transition rate from $\left|1 \right>$ to $\left|2 \right>$ due to two-photon reabsorption.  Since an atom will experience transition to many momentum states other than the $\left|1 \right>$ state once it is in the $\left|2 \right>$ state, population does not build up in $\left|2 \right>$, and the evolution from $\left|1 \right>$ to $\left|2 \right>$ is governed by a rate equation.  To calculate the rate, we input $\Omega_p$ and $\Omega_s$ into the optical Bloch equations for the three-state system shown in Fig.\ 1 \cite{opticalbloch}.  In the course of our calculation, we adiabatically eliminate the upper state, include transitions from $\left|1 \right>$ and $\left|2 \right>$ to other momentum states as phenomenological decay rates, ignore the effectively small AC stark shifts, and integrate the equation for the dipole ($\rho_{12}$ in standard notation) assuming slow variation of the populations of states $\left|1 \right>$ and $\left|2 \right>$.  Following this procedure, we find the transition rate ($R$),
\begin{eqnarray}
R \sim \{\frac{\Omega_1^4}{\Delta_1^2}+\frac{\Omega_2^4}{\Delta_2^2}+\frac{2\Omega_1^2\Omega2^2}{\Delta_1\Delta_2}\cos(\phi'-\theta')\}.
\end{eqnarray}
The dependence of the reabsorption rate on the presence of a spatial interference pattern is contained in the last term of Eq. (1).  In the absence of any interference pattern, the last term will contribute maximally to the reabsorption rate, as was the case with the second set of experiments.  With the presence of a rapidly varying interference pattern, $\theta'-\phi'$ will vary rapidly and the last term will average to zero over all of the atoms in the cloud and not contribute to the overall reabsorption rate.  The interference pattern created in the overlap of the two separate beams in the first set of experiments was sufficient to introduce enough phase variation to virtually eliminate the contribution due to the last term in Eq. (1), resulting in significantly less reabsorption.

For a physical interpretation of this result, we begin by noting that two-photon reabsorption involves a coherent transition.  If there were a sufficiently fast variation between the phase of the light driving the two legs of the transition, the transition rate would be reduced.  For the detunings we used, however, the response time of an atom's dipole ($\sim \frac{1}{\Delta}$) is a few nanoseconds.  Thus, for pump light with multiple frequency components but no interference pattern, the resulting spontaneous emission will follow the pump light effectively instantaneously.  There is no chance to introduce any variation between the phases of the pump and spontaneous emission light.  With an interference pattern, however, the source of the spontaneous emission light seen by an atom will vary as different sets of atoms are illuminated at different times.  Since there is no correlation in the positions of the atoms, this variation in which atoms are illuminated will in turn lead to a phase variation between the pump and spontaneous emission light, disrupting reabsorption.

\begin{figure}
\includegraphics{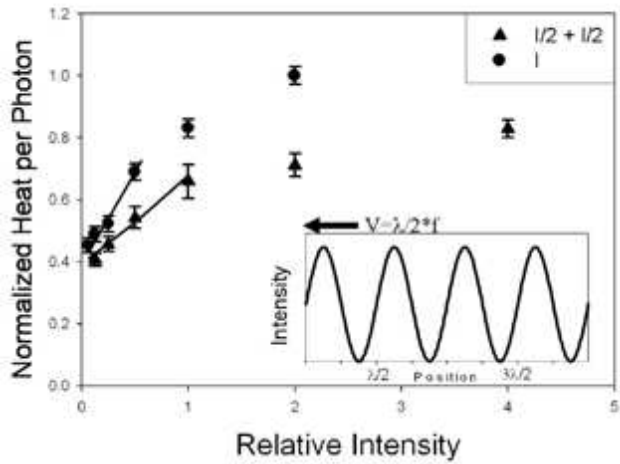}
\caption{\label{label}Heat per photon vs intensity for the single laser, counter-propagating experiment.  Circles are a single beam at full intensity, triangles are two simultaneously counter-propagating beams each at half intensity.  The slope gives the value of the $C_2$ coefficient.  The insert shows the interference pattern, which moves at velocity $v$.  $\lambda=780$ nm.}
\end{figure}
To further confirm our model's prediction, we revisited the two-frequency AOM experiments.  Again, two beams were created and then separated after the AOM.  However, unlike the previous experiments which used co-propagating beams, this time the heating pulses were made to be counter-propagating, resulting in an interference pattern (see Fig.\ 4).  This interference pattern meets all of the necessary conditions for reabsorption mitigation contained in our model.  Fig.\ 4 shows the results of one of these experiments.  Since we are plotting heat per photon ($C_{1}$+$C_{2}I$), the slope of the intensity profile before saturation occurs gives the coefficient for the two-photon Raman scattering process \cite{endnote}.  The intensity of the two separate heating pulses was adjusted so that each beam introduced the same amount of heat into the cloud when applied individually.  Our theory predicts for this set of conditions a reduction in $C_2$ by a factor of $\frac{1}{2}$; we measured a reduction of 0.457$\pm$.034.  Other tests with different parameters indicated the same general result \cite{jitter}.

Our semi-classical model of reabsorption predicts that the reduction in reabsorption scales linearly with the number of beams that have sufficient spatial and frequency variation from one another.  This is supported by the counter-propagating reabsorption test where we saw a reduction in the reabsorption by half.  Therefore, it should be possible to reduce the reabsorption beyond a factor of two by creating additional frequency components with the AOM, separating the components, and then either deliberately altering the phase of their beam profiles relative to one another with diffusers or sending them into the atom cloud at sufficiently different propagation directions.  Our current experimental configuration prevents such a test from being conducted in a practical manner, since the use of a free space MOT requires the heat beam path and imaging path to be the same to avoid problematic center-of-mass motion.  This limits the propagation directions and prevents the insertion of optics like diffusers in the heat beam paths.  However, in the future, the MOT atoms will be loaded into a far off resonance optical trap which will remove the restrictions encountered with a free space MOT, allowing tests with more than two frequencies to be conducted.

In summary, our experiments have successfully demonstrated the mitigation of reabsorption using beam modulation techniques.  Our results are explained by a semi-classical model.  The model indicates that the physics of multiple light scatterers in a random gas medium must be considered in order to understand our results.  Understanding the reabsorption process will allow current optical techniques to be modified, which should increase the efficiency of optical pumping schemes and promises to improve the lowest temperatures that can be achieved in optically thick gases using laser cooling techniques.

One of us (DMF) acknowledges support of the Research Corporation.  This work was funded by the AFOSR, grant number FA9550-06-1-0190.


\begin{thebibliography}{16}
\bibitem{photons}  For the purposes of this paper, pump photons are photons incident on the cloud, coming directly from the lasers.  Scattered photons are spontaneously emitted by an atom after it absorbs a pump photon.
\bibitem{kappanote} In writing the constant $\kappa$, we ignore the $\frac{1}{r}$ decrease in the amplitude of the spontaneously emission.  This approximation does not affect the model's relevant conclusions.
\bibitem{endnote}  These data were taken at lower intensities from previous sets, so that saturation was not a concern.
\bibitem{jitter}  This last set of experiments shows that the phase jitter between the separate lasers in the first set of experiments was not responsible for the reabsorption reduction.  Also, the first set of experiments shows that it is the interference pattern and not the change to the counter-propagating beams responsible for the reduction of reabsorption in this last set of data.
\end{thebibliography}
\end{document}